\documentstyle[tighten,epsf,floats,aps,preprint]{revtex}
\renewcommand{\vec}{\bf}

\preprint{RIKEN-AF-NP-214}
\title{Statistical properties of antisymmetrized molecular dynamics
for \\non-nucleon-emission and nucleon-emission processes}
\author{Akira Ono}
\address{Institute of Physical and Chemical Research (RIKEN),
Wako, Saitama 351-01, Japan}
\author{Hisashi Horiuchi}
\address{Department of physics, Kyoto University, Kyoto 606-01, Japan}

\begin{document}
\maketitle
\begin{abstract}
Statistical properties of the antisymmetrized molecular dynamics (AMD)
are classical in the case of nucleon-emission processes, while they
are quantum mechanical for the processes without nucleon emission.  In
order to understand this situation, we first clarify that there
coexist mutually opposite two statistics in the AMD framework: One is
the classical statistics of the motion of wave packet centroids and
the other is the quantum statistics of the motion of wave packets
which is described by the AMD wave function.  We prove the classical
statistics of wave packet centroids by using the framework of the
microcanonical ensemble of the nuclear system with realistic effective
two-nucleon interaction.  We show that the relation between the
classical statistics of wave packet centroids and the quantum
statistics of wave packets can be obtained by taking into account the
effects of the wave packet spread.  This relation clarifies how the
quantum statistics of wave packets emerges from the classical
statistics of wave packet centroids.  It is emphasized that the
temperature of the classical statistics of wave packet centroids is
different from the temperature of the quantum statistics of wave
packets.  We then explain that the statistical properties of AMD for
nucleon-emission processes are classical because nucleon-emission
processes in AMD are described by the motion of wave packet centroids.
We further show that when we improve the description of the
nucleon-emission process so as to take into account the momentum
fluctuation due to the wave packet spread, the AMD statistical
properties for nucleon-emission processes change drastically into
quantum statistics.  Our study of nucleon-emission processes can be
conversely regarded as giving another kind of proof of the fact that
the statistics of wave packets is quantum mechanical while that of
wave packet centroids is classical.
\end{abstract}
\pacs{25.40.Ep,02.70.Ns}

\narrowtext
\section{INTRODUCTION}

In order to have better description of fragmentation processes in
nuclear reactions, several molecular dynamics models with nucleon wave
packets have been developed in nuclear physics besides the molecular
dynamics with point nucleons imported from other fields such as
molecular physics and solid state physics.  An important question
about the wave packet molecular dynamics is whether its statistical
properties can be quantum mechanical unlike the point particle
molecular dynamics or not.  To have proper quantum statistics is
highly desirable for the molecular dynamics description of particle
emission and fragmentation processes in relatively low temperature
stage of nuclear reactions.

The purpose of this paper is to clarify in detail the statistical
properties of the antisymmetrized molecular dynamics
(AMD)\cite{ONO,ONOA} in general situations including both
non-nucleon-emission and nucleon-emission processes.  Many of our
arguments here can be equally applied to other kinds of wave packet
molecular dynamics. About the statistical properties of AMD, there
have been proposed two opposite opinions, one proposed by Ohnishi and
Randrup\cite{OHNISHI} who say that the AMD statistics is classical and
the other proposed Schnack and Feldmeier\cite{SCHNACK} who insist that
the AMD statistics is quantum mechanical as well as the statistics of
the fermionic molecular dynamics (FMD)\cite{FELD}.  Both two groups of
authors analysed the many-fermion system under harmonic oscillator
mean field when they deduce their conclusions.  Under this
controversial situation about AMD statistics, the present authors have
written recently a short paper\cite{ONOB} in which they have insisted
that the AMD statistics is quantum mechanical for non-nucleon-emission
processes while it is classical for nucleon-emission processes. In the
paper the present authors have also shown that an improved treatment
of nucleon-emission processes which takes into account the momentum
fluctuation due to the wave packet spread converts drastically the AMD
statistics for nucleon-emission processes from classical statistics
into quantum statistics.

Our present paper aims to give clear arguments which resolve the
above-mentioned controversial situation between Ref.\ \cite{OHNISHI}
and Ref.\ \cite{SCHNACK} and then based on this clarification it aims
to make the meaning of our previous paper of Ref.\ \cite{ONOB} more
understandable and the arguments of it more convincing for the
readers.

One of the basic points of this paper is to elucidate a unique feature
of AMD that two opposite statistics are coexistent: One is the
classical statistics which the centroids of nucleon wave packets obey
and the other is the quantum statistics which nucleon wave packets
obey in their motion described by the AMD wave function.  A key point
to understand this coexistent feature of different statistics is to
recognize that the temperature of the classical statistics of wave
packet centroids is different from the temperature which characterizes
the quantum statistics of wave packets. The fact that the motion of
wave packet centroids obeys classical statistics was proved by Ohnishi
and Randrup\cite{OHNISHI} by using the canonical ensemble technique in
simple cases of many-fermion system under harmonic oscillator mean
field and also in free gas state.  In this paper, we prove this fact
by using the microcanonical ensemble technique in the case of
realistic many-nucleon system interacting with effective two-nucleon
force.  The fact that the motion of wave packets described by the AMD
wave function obeys the quantum statistics was proved by Schnack and
Feldmeier\cite{SCHNACK} also in the case of harmonic oscillator
many-fermion system.  Their proof was made by calculating the
occupation probability of the harmonic oscillator single-particle
level by performing the long-time average in AMD (or FMD) and then by
showing that the occupation probability calculated with AMD (or FMD)
has just the same value as that given by the quantum mechanical
canonical ensemble average.  In this paper we prove this fact in
another way so that we can understand the relation between the
classical statistics of wave packet centroids and the quantum
statistics of wave packets.  Namely we will show that, when we
calculate the occupation probability of the harmonic oscillator
single-particle level on the basis of the fact that wave packet
centroids obey classical statistics, the resulting value of the
occupation probability proves to be almost the same as the value given
by the quantum mechanical canonical ensemble average.  Through this
proof we will see that the transition from the classical statistics of
wave packet centroids to the quantum statistics of wave packets is
surely due to effects of the wave packet spread.

In the case of the harmonic oscillator many-fermion system which both
Ohnishi-Randrup and Schnack-Feldmeier groups treated, the AMD
statistics should be regarded as being quantum mechanical just in
accordance with the opinion of Schnack and Feldmeier.  It is because
the AMD statistics that is relevant for calculated observables is not
the statistics of wave packet centroids but that of wave packets whose
motion is described by the AMD wave function.  We should note that the
calculation of observables is made by the use of the AMD wave
function.  We can say that the AMD statistics in the case of the
realistic nuclear system is also quantum mechanical so long as we are
concerned with dynamical processes confined spatially inside the
nucleus like the harmonic oscillator many-fermion system.

However, as we discussed in our previous paper\cite{ONOB}, the
situation changes when we treat nucleon-emission processes with AMD.
The nucleon-emission process can not be studied with the harmonic
oscillator many-fermion model nor the free gas model.  The AMD
statistics which is contained in nucleon-emission processes is now not
the statistics of wave packets but the statistics of the wave packet
centroids and hence it is classical.  The reason is because the
nucleon emission process in AMD is governed by the motion of the wave
packet centroid.  In AMD, a nucleon cannot go out from a nucleus if
the wave packet centroid cannot go out from the nucleus even when the
high momentum component of the wave packet can go out from the nucleus
if treated quantum mechanically.  In Ref.\ \cite{ONOB} we further
showed that we can recover the quantum mechanical character of the AMD
statistics also for nucleon-emission processes by introducing a
modified treatment of the nucleon-emission process. An essential point
of the modification is that we allow a wave packet near the nuclear
surface to split into high and low momentum components and allow the
high momentum component to escape from the nucleus.  Namely we take
into account dynamical effects of the momentum fluctuation due to the
wave packet spread.  In this paper we give a detailed discussion of
this modification on the basis of our clarified recognition of the
coexistence of mutually opposite two statistics in the AMD
framework. Our study of nucleon-emission processes with the ordinary
AMD and with the modified AMD can be conversely used as another kind
of proof of the fact that the statistics of wave packets is quantum
mechanical while that of wave packet centroids is classical.

The organization of this paper is as follows.  In the next section
(Sec.\ II) we explain briefly the AMD formalism.  Then in Sec.\ III,
we prove that the statistical properties of wave packet centroids are
classical.  The proof is performed by using the microcanonical
ensemble technique in the case of realistic many-nucleon system
interacting with effective two-nucleon force.  In Sec.\ IV we prove
that the statistical properties of wave packets whose motion is
described by the AMD wave function are quantum mechanical.  The proof
is made in the case of the harmonic oscillator many-fermion system by
calculating numerically the occupation probability of the harmonic
oscillator single-particle level on the basis of the fact that wave
packet centroids obey the classical statistics.  We will see that the
resulting value of the occupation probability proves to be almost the
same as the value given by the quantum mechanical canonical ensemble
average.  Here we further show that in the case of the harmonic
oscillator system of distinguishable many particles the proof can be
made totally analytically.  In Sec.\ V we discuss the statistical
properties of AMD for nucleon-emission processes. We then discuss a
modified treatment of the nucleon-emission process which recovers the
quantum mechanical character of the AMD statistics also for the
nucleon-emission process.  Summarizing discussions are given in Sec.\
VI.

\section{BRIEF EXPLANATION OF AMD FORMALISM}

We here explain briefly the AMD formalism.  For details the reader is
referred to Ref.\ \cite{ONO}.  AMD describes the nuclear many-body
system by a Slater determinant of Gaussian wave packets as
\begin{eqnarray}
\Phi(Z) &=& \det\Bigl[\phi({\vec r}_j,
 {\vec Z}_k)\chi_{\alpha_k}(j)\Bigr],
\nonumber\\
\phi({\vec r}, {\vec Z}) &=&  \Bigl(\frac{2\nu}{\pi}\Bigr)^{3/4}
\exp\Bigl\{-\nu\bigl({\vec r}-{\vec Z}/\sqrt{\nu}\bigr)^2
+\frac12{\vec Z}^2\Bigr\},
\end{eqnarray}
where the complex parameters $Z \equiv \{{\vec Z}_k\}$ stand for the
centroids of wave packets and $\chi_{\alpha}(j)$ for the spin-isospin
wave function with $\alpha = {\rm p}\uparrow, {\rm p}\downarrow, {\rm
n}\uparrow, {\rm n}\downarrow$.  The width parameter $\nu$ is common
to all nucleons and is time-independent.  The time evolution of $Z$ is
determined by the time-dependent variational principle and the
stochastic two-nucleon collision process.  The AMD equation of motion
for $Z$ derived from the time-dependent variational principle is
\begin{eqnarray}
&& i\hbar \sum_{k\tau} C_{j\sigma,k\tau} \frac {d Z_{k\tau}}
{d t} = \frac {\partial {\cal H}} {\partial Z^*_{j\sigma}},
\nonumber\\
&& C_{j\sigma,k\tau} = \frac {\partial^2} {\partial Z^*_{j\sigma}
\partial Z_{k\tau}} \log  \langle \Phi(Z)|\Phi(Z) \rangle,
\label{amdeq}
\end{eqnarray}
where $\sigma, \tau = x, y, z$ and the Hamiltonian function ${\cal H}$
is the expectation value of the Hamiltonian operator $H$ with the
subtraction of the spurious kinetic energies of the center-of-mass
zero-point oscillations of fragments,
\begin{equation}
{\cal H} = \frac { \langle \Phi(Z)|H|\Phi(Z) \rangle }
{\langle \Phi(Z)|\Phi(Z) \rangle} - \frac {3\hbar^2\nu} {2M} A
+ T_0 ( A - N_F ).
\label{amdhamil}
\end{equation}
Here $A$ is the total mass number, $M$ is the nucleon mass, $N_F$ is
the fragment number which is a function of $Z$ and $Z^*$, and $T_0$ is
${3\hbar^2\nu}/ {2M}$ in principle but is treated usually as a free
parameter for the adjustment of binding energies.

For the stochastic two-nucleon process, we introduce physical nucleon
coordinates $W = \{{\vec W}_j\}$ as
\begin{eqnarray}
{\vec W}_j &=& \sum_{k=1}^A \Bigl(\sqrt Q \Bigr)_{jk} {\vec Z}_k,
\nonumber\\
Q_{jk} &=&  \frac \partial {\partial ({\vec Z}^*_j \cdot {\vec Z}_k)}
\log \langle \Phi(Z)|\Phi(Z) \rangle.
\label{physco}
\end{eqnarray}
In the case of light systems where we have less than three nucleons in
any spin-isospin state, the physical nucleon coordinates $W$ are
canonical coordinates.  However in other cases $W$ are not canonical
coordinates.  An important property of $W$ is that in the many-body
phase space of $W$ there exists a Pauli-forbidden region into which
any nucleon cannot enter.  For example, we can easily show that the
following region of $W$ is contained in the Pauli-forbidden region,
\begin{equation}
\sum_{j=1}^A {{\vec W}^*_j} \cdot {{\vec W}_j} < N_A,
\end{equation}
where $N_A$ is the minimum number of the total harmonic oscillator
quanta of the $A$-nucleon system.  The Pauli blocking of the
two-nucleon collision is defined such that the two-nucleon collision
should avoid entering into the Pauli-forbidden region.

The AMD equation of motion (Eq.\ (\ref{amdeq})) is not of canonical
form, indicating that the coordinates $Z$ are not canonical
coordinates. Although the existence of canonical coordinates is known
mathematically, we do not know at present the explicit forms of
canonical coordinates in AMD except in the case of light systems
mentioned above (just below Eq.\ (\ref{physco})).  However, even if we
do not know the explicit forms of canonical coordinates, we can derive
the explicit form of the Jacobian of the transformation between
canonical coordinates and the original coordinates $Z$.  If we express
a set of canonical coordinates by $Y$ = $\{ {\vec Y}_j \}$, the
Jacobian $J$ can be shown to be given by the determinant of the
hermitian matrix $\{C_{j\sigma,k\tau}\}$ appearing in the AMD equation
of motion (Eq.\ (\ref{amdeq})),
\begin{equation}
J = \frac {\partial (Y,Y^*)} {\partial (Z,Z^*)} =
\frac {\partial (R,P)} {\partial (D,K)} =
\det \bigl[C_{j\sigma,k\tau}\bigr],
\label{jacobian}
\end{equation}
where $R$ = $\{{\vec R}_j\}$ and $P$ = $\{{\vec P}_j\}$ are real and
imaginary parts of $Y$, respectively, while $D$ = $\{{\vec D}_j\}$ and
$K$ = $\{{\vec K}_j\}$ are those of $Z$:
\begin{equation}
{\vec Y}_j = {\sqrt \nu} {\vec R}_j
 + \frac i {2\hbar {\sqrt \nu}} {\vec P}_j,  \quad
{\vec Z}_j = {\sqrt \nu} {\vec D}_j
 + \frac i {2\hbar {\sqrt \nu}} {\vec K}_j.
\end{equation}
In spite of our general ignorance about explicit forms of canonical
coordinates, there exists one canonical coordinate whose explicit form
we know.  It is the center of mass coordinate ${\vec Z}_G$,
\begin{eqnarray}
{\vec Z}_G &=& \frac 1 {\sqrt A} \sum_{j=1}^A {\vec Z}_j
= \sqrt {A \nu} {\vec D}_G + \frac i {2\hbar \sqrt {A \nu}}
{\vec K}_G, \nonumber \\
{\vec D}_G &=& \frac 1 A \sum_{j=1}^A {\vec D}_j, \quad
{\vec K}_G = \sum_{j=1}^A {\vec K}_j.
\label{cmc}
\end{eqnarray}

For the sake of discussions in later sections, we here emphasize the
fact that the AMD wave function is just the exact solution of the
time-dependent Schr\"odinger equation in the case of the harmonic
oscillator mean field.  Namely, when the Hamiltonian operator $H$ is
given as
\begin{equation}
H = \hbar \omega \sum_{j=1}^A {\vec a}^\dagger_j \cdot {\vec a}_j
 + E_0,
\end{equation}
where ${\vec a}^\dagger_j$ and ${\vec a}_j$ ($j = 1 \sim A$) are
harmonic oscillator creation and annihilation operators, the exact
solution of the Schr\"odinger equation $i\hbar ( \partial / \partial t
) \Psi$ = $H \Psi$ is given as
\begin{equation}
\Psi = e^{-i E_0 t / \hbar}
\det \Bigl[ \phi ( {\vec r}_j, e^{-i\omega t}
{\vec Z}_k^{(0)} ) \chi_{\alpha_k}(j) \Bigr].
\end{equation}
Here the width parameter $\nu$ of the Gaussian wave packet $\phi$ is
$\nu$ = $M\omega / 2\hbar$, and the parameters $\{{\vec Z}_k^{(0)}\}$
stand for arbitrary initial values for $Z$ = $\{{\vec Z}_k\}$.
Needless to say, $\{{\vec Z}_k$ = $e^{-i\omega t} {\vec Z}_k^{(0)}, (
k=1 \sim A )\}$ are the solution of the AMD equation of motion (Eq.\
(\ref{amdeq})) with $T_0 = 3\hbar \nu /2M$.  We should note the fact
that the AMD wave function which is the exact solution of the
Schr\"odinger equation is constructed from the motion of wave packet
centroids which make just the classical oscillatory motion $\{{\vec
Z}_k = e^{-i\omega t} {\vec Z}_k^{(0)}, ( k=1 \sim A )\}$.  This fact
constitutes the basic background of the character of the AMD
statistical properties which we clarify in later sections.  There we
will see that the quantum statistics of AMD is constructed from the
statistics of wave packet centroids which is classical.

\section{CLASSICAL STATISTICS OF WAVE PACKET CENTROIDS}

We here prove the fact that the motion of wave packet centroids in AMD
obeys classical statistics.  The statistical properties of wave packet
centroids are determined by the AMD equation of motion.  We consider
that stochastic two-nucleon collisions are not relevant to the
character of the statistics but play only the role to hasten the
equilibration of the system.  The AMD equation of motion of Eq.\
(\ref{amdeq}) is not of canonical form but it can be converted into
the canonical form if we use canonical coordinates instead of $Z$.
Hence the dynamics of wave packet centroids is classical and the
statistical properties of wave packet centroids are expected to be
classical.  However, what inhibits us from concluding
straightforwardly in this way is the effect of the antisymmetrization
contained in the AMD equation of motion.  As we explained in Sec.\ II,
there exists a Pauli-forbidden region in the phase space of the
physical coordinates $W$.  Although the physical coordinates $W$ are
not always equivalent to canonical coordinates, we can guess the
existence of a Pauli-forbidden region also in the phase space of
canonical coordinates.  If we have a Pauli-forbidden region, we are
not sure whether we can simply say or not that the statistics of wave
packet centroids is classical.  Under this situation, we are forced to
check explicitly the character of the statistics of wave packet
centroids.

In Ref.\ \cite{OHNISHI}, Ohnishi and Randrup investigated the
character of the statistics of wave packet centroids by using the
canonical ensemble technique in simple cases of many-fermion system
under one-dimensional harmonic oscillator mean field and also in free
gas state.  The result they obtained is that the statistics of wave
packet centroids is classical.  Here in this paper, we investigate the
the character of the statistics of wave packet centroids by using the
microcanonical ensemble technique in the case of realistic
many-nucleon system interacting with effective two-nucleon force.

In the microcanonical ensemble method, the relation between the energy
$E$ and the temperature $T_c$ is given by
\begin{eqnarray}
\label{micro}
&&\frac 1 {T_c} = \frac {\partial \log \rho (E)} {\partial E},
\nonumber\\
\rho (E) &&=  \int \Bigl( \prod_{j=1}^A \frac {d^3R_j d^3P_j}
{(2\pi \hbar)^3}
\Bigr) \delta ( {\cal H} - E ) \delta ( {\vec D}_G )
\delta ( {\vec K}_G )\\
=&& \int \Bigl( \prod_{j=1}^A \frac {d^3D_j d^3K_j} {(2\pi \hbar)^3}
\Bigr) \det \bigl[ C \bigr] \delta ( {\cal H} - E )
\delta ( {\vec D}_G ) \delta ( {\vec K}_G )  .
\nonumber
\end{eqnarray}
Here $\det [C] = \det \bigl[ C_{j\sigma,k\tau} \bigr]$ and we have
used the relation of Eq.\ (\ref{jacobian}) for the Jacobian between
$Z$ and the canonical coordinates $Y$. The existence of $\delta (
{\vec D}_G ) \delta ( {\vec K}_G )$ is because of the fact that the
center of mass coordinate ${\vec Z}_G$ (Eq.\ (\ref{cmc})) is a
conserved quantity and makes no contribution to the phase volume $\rho
(E)$ of the internal motion of the nucleus.  For the sake of the
numerical evaluation of this relation, we rewrite this relation as
explained below\cite{PAULI}. We first notice the following identity
relation
\widetext
\begin{eqnarray}
& & \frac {\partial} {\partial E} \int  dZ \det [C]
\delta ( {\cal H} - E ) \delta ( {\vec D}_G ) \delta ( {\vec K}_G )
\sum_{\alpha=1}^{6A} a_\alpha
\frac {\partial {\cal H} } {\partial z_\alpha}
\nonumber\\
& & =  \int dZ \det [C]
\delta ( {\cal H} - E ) \delta ( {\vec D}_G ) \delta ( {\vec K}_G )
\sum_{\alpha=1}^{6A} \Bigl(
{\frac {\partial a_\alpha} {\partial z_\alpha}} +
a_\alpha {\frac {\partial \log \det [C] } {\partial z_\alpha}} +
a_\alpha {\frac {\partial \log \delta ( {\vec D}_G )
\delta ( {\vec K}_G )} {\partial z_\alpha}} \Bigr),
\label{identity}
\end{eqnarray}
\narrowtext\noindent
where
\begin{eqnarray}
dZ &=&  \prod_{j=1}^A \frac {d^3D_j d^3K_j} {(2\pi \hbar)^3},
\nonumber\\
\{ z_\alpha, \alpha = 1 \sim 6A \} &=&
\{ {\vec D}_j, {\vec K}_j, j = 1 \sim A \},
\end{eqnarray}
and $\{ a_\alpha \}$ are arbitrary functions of $Z$.
The identity relation of Eq.\ (\ref{identity}) can be easily
derived by using $( \partial / \partial E \delta ( {\cal H} - E ) )
( \partial {\cal H} / \partial z_\alpha ) = - \partial /
\partial z_\alpha \delta ( {\cal H} - E )$ and integration
by parts.  We choose $\{ a_\alpha \}$ so that they satisfy
\begin{equation}
\sum_{\alpha} a_{\alpha} \frac {\partial {\vec D}_G}
{\partial z_\alpha} = 0, \quad
 \sum_{\alpha} a_{\alpha} \frac {\partial {\vec K}_G}
{\partial z_\alpha} = 0.
\label{cmcadjust}
\end{equation}
For those $\{ a_\alpha \}$ which satisfy Eq.\ (\ref{cmcadjust}), we
have
\begin{equation}
\sum_{\alpha} a_\alpha {\frac {\partial \log \delta ( {\vec D}_G )
\delta ( {\vec K}_G )} {\partial z_\alpha}} = 0.
\end{equation}
By dividing both sides of the identity relation Eq.\ (\ref{identity})
by $\rho ( E )$ and by using Eq.\ (\ref{micro}), we get
\begin{equation}
T_c = \frac {\displaystyle \langle \sum_\alpha a_\alpha
{\frac {\partial {\cal H}}
{\partial z_\alpha} } \rangle} { \displaystyle
\langle \sum_\alpha \Bigl(
{\frac {\partial a_\alpha} {\partial z_\alpha}} +
a_\alpha {\frac {\partial \log \det [C] } {\partial z_\alpha}}
\Bigr) \rangle - {\frac \partial {\partial E}}
\langle \sum_\alpha a_\alpha {\frac {\partial {\cal H}}
{\partial z_\alpha} } \rangle}  ,
\label{microtem}
\end{equation}
where $\langle Q(Z) \rangle$ stands for the microcanonical ensemble
average of the quantity $Q(Z)$ for the energy $E$ and is defined as
\begin{equation}
\langle Q(Z) \rangle = \frac {\displaystyle \int dZ
\det [C] \delta ( {\cal H} - E ) \delta ( {\vec D}_G )
\delta ( {\vec K}_G ) Q(Z)} {\displaystyle \rho ( E )}.
\end{equation}

Under the thermally equilibrated situation of the system, we can
equate the microcanonical ensemble average $\langle Q(Z) \rangle$ to
the long-time average $\overline {Q(Z)}$;
\begin{eqnarray}
\langle Q(Z) \rangle &=& \overline {Q(Z)},
\nonumber\\
\overline {Q(Z)} &=& \lim_{t_2 \rightarrow \infty}
\frac 1 {t_2 - t_1} \int_{t_1}^{t_2} dt Q(Z(t)).
\end{eqnarray}
By using this equivalence of $\langle Q(Z) \rangle$ with $\overline
{Q(Z)}$, we can numerically evaluate Eq.\ (\ref{microtem}) by
calculating the time development of $Z$ with the AMD equation of
motion for a long time. We give in Fig.\ 1 the calculated relation
between the temperature $T_c$ and the energy $E$ in the system of
$^{12}$C. Since the contribution of the term $(\partial/\partial E)
\langle \sum_\alpha a_\alpha \partial {\cal H}/
\partial z_\alpha \rangle$
in the denominator of Eq.\ (\ref{microtem}) is very small, we
neglected this term in the calculation. The arbitrary functions $\{
a_\alpha \}$ were here selected to be
\begin{eqnarray}
a_\alpha &=& z_\alpha - \frac 1 A \sum_{k=1}^A
D_{k \sigma} \quad \quad {\rm for} \quad
z_\alpha = D_{j \sigma}, \nonumber \\
a_\alpha &=& z_\alpha - \frac 1 A \sum_{k=1}^A
K_{k \sigma} \quad \quad {\rm for} \quad
z_\alpha = K_{j \sigma}.
\end{eqnarray}
We can easily check that these $\{ a_\alpha \}$ surely satisfy Eq.\
(\ref{cmcadjust}). We adopted the Volkov No.1 force\cite{VOLKOV} for
the effective two-nucleon force.  The oscillator width $\nu$ was
chosen to be $\nu$ = 0.16 fm$^{-2}$.  It is easy to see that the
relation between the excitation energy $E^* = E - E({\rm ground\
state})$ and the temperature $T_c$ is very close to the caloric curve
$E^* / 11 = 3 T_c$.  Since this means that the specific heat remains
constant even around zero temperature, the statistics is not quantum
mechanical.  It is well known that the caloric curve $E^* / A = 3 T_c$
is characteristic to the classical statistics of the harmonic
oscillator many-body system. We can say that the reason why the
caloric curve in Fig.\ 1 is closer to $E^* / 11 = 3 T_c$ than $E^* /
12 = 3 T_c$ is because in our present approach by the use of the
microcanonical ensemble, the center of mass coordinate is not involved
in the thermal motion implying that the number of degrees of freedom
is $3(A-1)$ rather than $3A$.  We thus conclude that the statistics of
AMD wave packet centroids for the realistic nuclear system is
classical.

\begin{figure}
\ifx\epsfxsize\undefined\else
\begin{minipage}[b]{0.5\textwidth}
\epsfxsize\textwidth\epsfbox{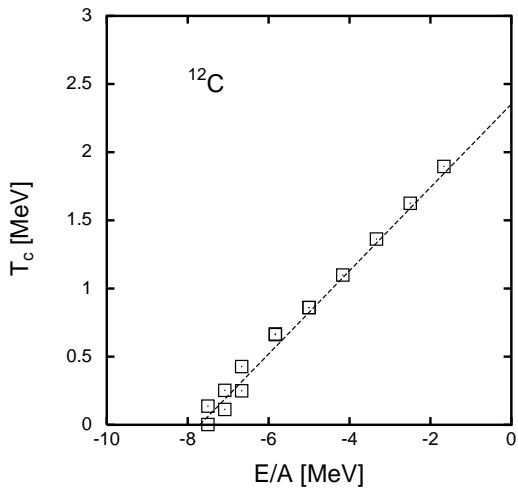}
\end{minipage}
\fi
\begin{minipage}[b]{0.48\textwidth}
\caption{The relation between the temperature $T_c$ and the energy
$E$ for the system of ${}^{12}{\rm C}$. Squares are the calculated
results with the microcanonical ensemble method. Different values of
$T_c$ for the same $E$ is due to the statistical error. Dotted line
shows the relation $E^*/(A-1)=3T_c$ with $A=12$.}
\end{minipage}
\end{figure}

\section{QUANTUM STATISTICS OF WAVE PACKETS}

The fact that the motion of wave packets described by the AMD wave
function obeys the quantum statistics was first given an explicit
proof by Schnack and Feldmeier\cite{SCHNACK}.  They treated the same
model system as Ohnishi and Randrup\cite{OHNISHI}, namely the
one-dimensional harmonic oscillator many-fermion system.  They also
treated the harmonic oscillator system of distinguishable particles
and found also the quantum mechanical character for the statistics of
wave packets.  Their proof was made by calculating the occupation
probability of the harmonic oscillator single-particle level by
performing the long-time average in AMD (or FMD) and then by showing
that the occupation probability calculated with AMD (or FMD) has just
the same value as that given by the quantum mechanical canonical
ensemble average.  Here in this paper, we prove the quantum statistics
of wave packets in another way so that we can understand its relation
with the classical statistics of wave packet centroids.  Our proof is
made by treating the same one-dimensional harmonic oscillator systems
of fermions and also of distinguishable particles.

Like Schnack and Feldmeier, we investigate the occupation probability
of the harmonic oscillator single-particle level.  In AMD, this
quantity which we express as ${\cal P}_n (Z)$ for the AMD wave
function $\Phi (Z)$ is given by
\begin{eqnarray}
& & {\cal P}_n (Z) =
\frac { \langle \Phi(Z)|P_n|\Phi(Z) \rangle }
{\langle \Phi(Z)|\Phi(Z) \rangle}, \nonumber \\
& & P_n = C_n^\dagger C_n = \sum_{j=1}^A |\phi_n(x_j)\rangle
\langle \phi_n(x_j)|,
\end{eqnarray}
where $\phi_n(x)$ is the one-dimensional harmonic oscillator
eigenfuncion with oscillator quanta $n$. Here the AMD wave function
$\Phi(Z)$ should be considered to be a Slater determinant of
one-dimensional Gaussian wave packet $\phi(x,Z_j) = \sqrt {2\nu/\pi}
\exp \{ -\nu (x - Z_j/{\sqrt \nu})^2 + (1/2)Z_j^2 \}$
without spin-isospin wave function (or belonging to a single
spin-isospin state). The investigation of ${\cal P}_n (Z)$ is made in
thermally equilibrated situation of the system. In this section we
describe the thermal equilibrium by the canonical ensemble method. It
means that we investigate the statistical ensemble of AMD wave
functions whose centroids $Z$ follow the classical canonical
statistics.  Let $T_c$ denote the temperature of the motion of wave
packet centroids.  The average value of ${\cal P}_n (Z)$ under thermal
equilibrium, which we express as $\langle {\cal P}_n (Z)
\rangle_{T_c}$, is calculated as
\begin{eqnarray}
\langle {\cal P}_n (Z) \rangle_{T_c} &=&
\frac {\displaystyle\int dZ \det [C] {\cal P}_n (Z)
e^{ - {\cal H}(Z)/ T_c} }
{\displaystyle
\int dZ \det [C] e^{ - {\cal H}(Z)/ T_c}}, \nonumber \\
{\cal H}(Z) &=& \frac {\langle \Phi(Z)|\hbar \omega \sum_{j=1}^A
(a^\dagger_j a_j + 1/2)|\Phi(Z) \rangle}
{\langle \Phi(Z)|\Phi(Z) \rangle}.
\end{eqnarray}
Here the notations, $dZ, \det [C],$ and others are obvious
one-dimensional analogues of the corresponding three-dimensional
quantities; for example,
\begin{eqnarray}
dZ &=&  \prod_{j=1}^A \frac {dD_j dK_j} {2\pi \hbar},
\nonumber\\
\det [C] &=&  \det \Bigl[\frac {\partial^2} {\partial Z^*_j
\partial Z_k} \log  \langle \Phi(Z)|\Phi(Z) \rangle  \Bigr].
\end{eqnarray}
Similarly, the average value of the energy under thermal equilibrium,
which we express as $\langle {\cal H}(Z) \rangle_{T_c}$, is calculated
as
\begin{equation}
\langle {\cal H}(Z) \rangle_{T_c} =
\frac {\displaystyle\int dZ \det [C] {\cal H}(Z)
e^{ - {\cal H}(Z)/ T_c} }
{\displaystyle\int dZ \det [C] e^{ - {\cal H}(Z)/ T_c}}.
\end{equation}

We have numerically calculated $\langle {\cal P}_n (Z) \rangle_{T_c}$
and $\langle {\cal H}(Z) \rangle_{T_c}$ by using the Metropolis
sampling method as in Ref.\ \cite{OHNISHI} in the case of the system
with fermion number 4 ($A$ = 4).  The quantity $\langle {\cal H}(Z)
\rangle_{T_c}$ is just the quantity which Ohnishi and Randrup
calculated in Ref.\ \cite{OHNISHI}.  We have ascertained their result
in the case of $A$ = 4; namely the calculated value of $\langle {\cal
H}(Z) \rangle_{T_c}$ is given almost exactly as
\begin{equation}
\langle {\cal H}(Z) \rangle_{T_c} = A T_c +
\hbar \omega \sum_{n=0}^{A-1} (n + \frac 1 2).
\end{equation}
We suspect that this relation is analytically exact, since our
numerical calculation shows that this equality holds with very high
accuracy. This result, of course, shows that the motion of wave packet
centroids obeys the classical statistics.

Now let us consider the quantity $\langle {\cal P}_n (Z)
\rangle_{T_c}$.  In classical mechanics, we have no such
quantity as $\langle {\cal P}_n (Z) \rangle_{T_c}$, because there is
no concept of the discrete energy level in classical mechanics.
Therefore we have no idea what value the quantity $\langle {\cal P}_n
(Z) \rangle_{T_c}$ should take in the case of classical statistics.
One may think that we have the relation $\langle {\cal P}_n (Z)
\rangle_{T_c} \propto \exp ( -n \hbar \omega / T_c )$ with some small
 modification by the Pauli principle.  However, we will soon see that
the $n$-dependence of the calculated values of $\langle {\cal P}_n (Z)
\rangle_{T_c}$ is very different from that of $\exp ( -n \hbar \omega
/ T_c )$.  Since the quantity $\langle {\cal P}_n (Z) \rangle_{T_c}$
is a quantum mechanical quantity, we should check whether the
calculated values of $\langle {\cal P}_n (Z)
\rangle_{T_c}$ are equal or not to what the quantum statistics gives
us.

In the quantum statistics, the average value of the occupation
probability of the harmonic oscillator single-particle level, which we
express as $\langle\langle P_n \rangle \rangle_T$, is given as
\begin{eqnarray}
\langle \langle P_n \rangle \rangle_T &=&
\frac {\displaystyle \sum_{n_1...n_4} {}^a\!\langle n_1...n_4 |
P_n | n_1...n_4 \rangle\!^a e^{ -E(n_1...n_4)/T }}
{\displaystyle \sum_{n_1...n_4} {}^a\!\langle n_1...n_4 |
n_1...n_4 \rangle\!^a e^{ -E(n_1...n_4)/T }}, \nonumber \\
|n_1...n_4 \rangle\!^a &=& \frac 1 {\sqrt {4!}} \det [ \phi_{n_1}
(x_1)...\phi_{n_4}(x_4)], \\
E(n_1...n_4) &=& \sum_{j=1}^4 ( n_j + \frac 1 2 )
\hbar \omega. \nonumber
\end{eqnarray}
Similarly, in the quantum statistics, the average value of
the energy, which we express as $\langle
\langle H \rangle \rangle_T$, is given as
\begin{equation}
\langle \langle H \rangle \rangle_T =
\frac {\displaystyle \sum_{\! n_1...n_4\!\!\!\!\!\!}
 {}^a\!\langle n_1...n_4 | n_1...n_4 \rangle\!^a E(n_1...n_4) e^{ -
E(n_1...n_4)/T }} {\displaystyle \sum_{n_1...n_4} {}^a\!\langle
n_1...n_4 | n_1...n_4 \rangle\!^a e^{ -E(n_1...n_4)/T }}.
\end{equation}

The necessary and sufficient condition that the system has quantum
statistical properties is that any of thermally averaged quantities
like $\langle {\cal P}_n(Z) \rangle_{T_c}$ and $\langle {\cal H}(Z)
\rangle_{T_c}$ are expressed by the thermally averaged quantities in
quantum canonical ensemble with a certain temperature $T$. What is
decisively important here is the recognition that there is no need
that the temperature $T$ of the quantum statistics is equal to the
temperature $T_c$ of wave packet centroids. The relation between the
temperatures $T$ and $T_c$ can be obtained for example by the
requirement that the value of $\langle {\cal H}(Z) \rangle_{T_c}$ is
equal to the value $\langle \langle H \rangle \rangle_T$.

We here summarize as follows what we are going to check now: We first
calculate numerically the relation between $T$ and $T_c$ from the
relation
\begin{equation}
\langle \langle H \rangle \rangle_T =
\langle {\cal H}(Z) \rangle_{T_c},
\end{equation}
Then we check whether the relation,
\begin{equation}
\langle \langle P_n \rangle \rangle_T =
\langle {\cal P}_n (Z) \rangle_{T_c},
\label{check}
\end{equation}
holds or not.  In Fig.\ 2 we show that the numerical calculations
assure very well the relation of Eq.\ (\ref{check}).  This result
demonstrates clearly that the statistics of the motion of wave packets
described by the AMD wave function is quantum mechanical.

\begin{figure}
\ifx\epsfxsize\undefined\else
\begin{center}
\epsfxsize0.8\textwidth~\epsfbox{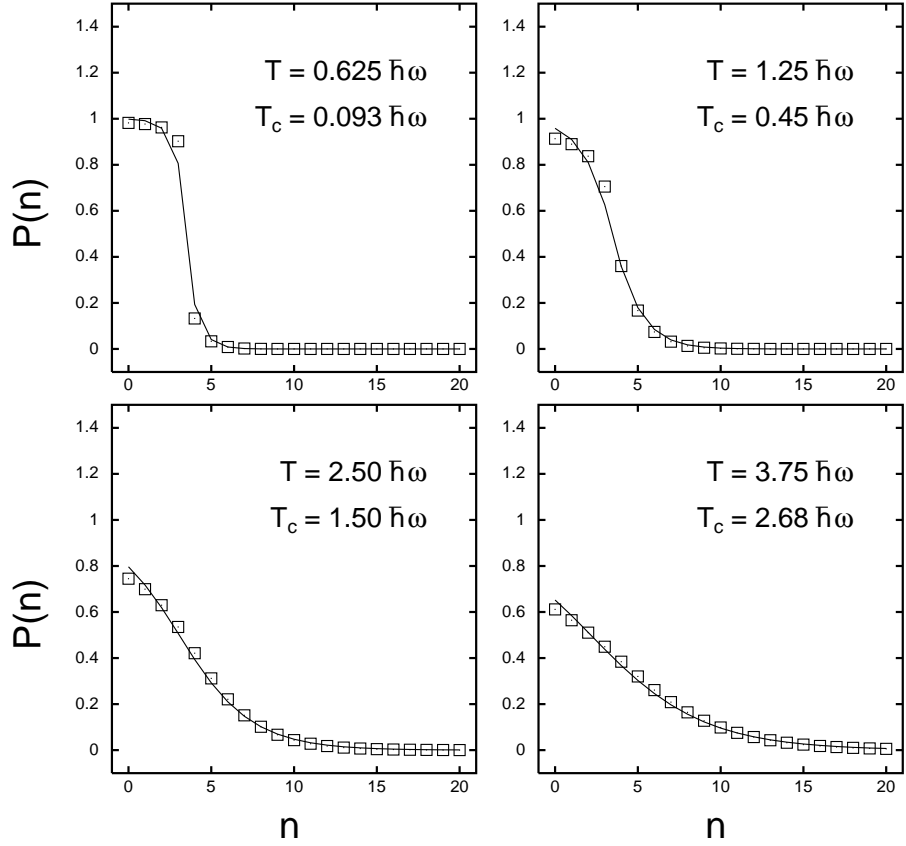}~
\end{center}
\fi
\caption{Occupation probability of the single-particle level
$n$ for the one-dimensional harmonic oscillator system with 4
fermions. Squares are the results of $\langle{\cal
P}_n(Z)\rangle_{T_c}$ calculated by assuming the classical canonical
statistics of the wave packet centroids. Lines show the quantum
mechanical relation $\langle\langle P_n\rangle\rangle_T$. $T$ and
$T_c$ of each panel are chosen so that they satisfy $\langle\langle
H\rangle\rangle_T=\langle{\cal H}(Z)\rangle_{T_c}$.}
\end{figure}

Our arguments given above in this section can be made completely
analytically in the case of distinguishable particles.  For
distinguishable particles, the AMD wave function is simply a product
of Gaussian wave packets,
\begin{equation}
\Phi(Z) = \prod_{j=1}^A \phi (x_j, Z_j).
\end{equation}
The coordinates $Z = \{ Z_j \}$ are now canonical coordinates and we
have $\det [C]$ = 1. The calculation of $\langle {\cal H}(Z)
\rangle_{T_c}$ can be executed analytically, giving the result
\begin{eqnarray}
{\cal H}(Z) &=& \sum_{j=1}^A ( Z_j^*Z_j + \frac 1 2 ) \hbar \omega,
\nonumber \\
\langle {\cal H}(Z) \rangle_{T_c} &=& A ( T_c + {\frac 1 2} \hbar
\omega ).
\end{eqnarray}
The calculated result of $\langle {\cal H}(Z) \rangle_{T_c}$ shows, of
course, the classical statistics of the motion of wave packet
centroids.  The calculation of $\langle \langle H \rangle \rangle_T$
for the quantum canonical ensemble can be easily made as shown in
ordinary text books of statistical mechanics and we have
\begin{equation}
\langle \langle H \rangle \rangle_T = A \Bigl(
\frac {\hbar \omega} {\exp (\hbar \omega / T) - 1}
+ \frac 1 2 \hbar \omega \Bigr).
\end{equation}
{}From the relation of $\langle \langle H \rangle \rangle_T$ =
$\langle {\cal H}(Z) \rangle_{T_c}$, we get the relation between $T$
and $T_c$ as follows
\begin{equation}
T_c = \frac {\hbar \omega} {\exp (\hbar \omega / T) - 1}.
\label{relation}
\end{equation}

The quantities $\langle {\cal P}_n (Z) \rangle_{T_c}$ and $\langle
\langle P_n \rangle \rangle_T$ can be also easily calculated
analytically, and we have
\begin{eqnarray}
&& {\cal P}_n (Z) = \sum_{j=1}^A \frac
{(Z_j^*Z_j)^n} {n!} \exp ( - Z_j^*Z_j ), \nonumber \\
&& \langle {\cal P}_n (Z) \rangle_{T_c} =
A \frac {\hbar \omega / T_c}
{( 1 + \hbar \omega / T_c)^{n+1}}, \\
&& \langle \langle P_n \rangle \rangle_T =
A \{ 1 - \exp ( - \hbar \omega / T ) \}
\exp ( -n \hbar \omega / T ). \nonumber
\end{eqnarray}
In getting the results for $\langle {\cal H}(Z) \rangle_{T_c}$ and
$\langle {\cal P}_n (Z)\rangle_{T_c}$, we have used the formula
%
%
\begin{equation}
\frac {\displaystyle
\int dD_j dK_j {\displaystyle \frac {(Z_j^*Z_j)^n} {n!}}
e^{-\alpha Z_j^*Z_j}
e^{-\beta Z_j^*Z_j} }
{\displaystyle
\int dD_j dK_j e^{-\beta Z_j^*Z_j}}
= \frac {\beta}
{( \alpha + \beta)^{n+1}}.
\label{formula}
\end{equation}
When we use the relation between $T$ and $T_c$ of Eq.\
(\ref{relation}), we can easily prove the equality
\begin{equation}
\langle {\cal P}_n (Z) \rangle_{T_c} =
\langle \langle P_n \rangle \rangle_T
\propto \exp ( -n \hbar \omega / T ).
\end{equation}
Thus we have proved the quantum statistics of the motion of wave
packets completely analytically in the case of distinguishable
particles under the harmonic oscillator mean field.  This proof in the
case of distinguishable particles clearly shows that the transition
from the classical statistics of wave packet centroids to the quantum
statistics of wave packets is due to effects of the wave packet
spread, because there is no complexity due to the antisymmetrization.

\section{AMD STATISTICS FOR THE NUCLEON-EMISSION PROCESS
AND INCORPORATION OF WAVE PACKET FLUCTUATION INTO AMD}

In previous sections we have explained a very unique feature of AMD
that in the AMD framework two opposite statistics are coexistent: One
is the classical statistics which the centroids of nucleon wave
packets obey and the other is the quantum statistics which nucleon
wave packets obey in their motion described by the AMD wave function.
We have pointed out that a key point to understand this coexistent
feature of different statistics is to recognize that the temperature
of the classical statistics of wave packet centroids is different from
the temperature of the quantum statistics of wave packets.

In AMD, the time development of the system is described only through
the time development of wave packet centroids which is calculated by
the AMD equation of motion and the stochastic two-nucleon collision.
The effects of the wave packet spread are taken into account through
the use of the AMD wave function for calculating observables.  This
ordinary AMD framework is expected to work well for the treatment of
the internal motion of the nucleus which is free from the nucleon
emission. It is because, in the case of the model system under the
harmonic oscillator mean field where we have no particle emission, the
AMD wave function is just the exact solution of the time-dependent
Schr\"odinger equation.  When the ordinary AMD framework is accepted
to work well for non-nucleon-emission processes, the AMD statistics
should be regarded as being not the classical statistics of wave
packet centroids but the quantum statistics of wave packets, because
observables are calculated by the use of the AMD wave function which
contains effects of the wave packet spread. The statistical ensemble
of AMD wave functions is expected to describe well quantum statistical
distributions of single-particle level (and/or momentum) occupation
probability.

However when we look at the description of the nucleon-emission
process by AMD, we find that the description only through the time
development of wave packet centroids is not always appropriate.  In
AMD, a nucleon can be emitted only when the wave packet centroid can
go out of the residual nucleus.  In quantum mechanics, however, even
if the wave packet centroid can not go out of the nucleus, the high
momentum component of the wave packet is allowed to go out of the
nucleus.  Namely, although the wave packet spread should take part
explicitly in the dynamics of the nucleon emission, the ordinary AMD
framework does not take account of this role of the wave packet
spread.  It should be noted that the wave packet spread in AMD is
large because the energy spread ($3\hbar^2\nu/2M$) due to the wave
packet spread is about 10 MeV.  The neglect of the nucleon emission
due to the high momentum component of the wave packet spread will
result, for example, in the small cross section of emitted nucleons or
the slowdown of nucleon emission process in the dynamical stage of the
reaction.  Of course for a high energy nucleon whose wave packet
centroid can easily escape from the nucleus, the ordinary AMD
description has no serious problem.

According to the above discussion, the emitted nucleons do not reflect
the correct momentum distribution inside the nucleus. They reflect
only the momentum distribution of wave packet centroids.  It means
that the statistical properties of the nucleon-emission process can
not be quantum mechanical.

In order to check the statistical properties of the nucleon-emission
process, we investigated in Ref.\ \cite{ONOB} the equilibrated
coexistent situation of gas and liquid nucleons.  For the sake of
self-contained discussion, we here recapitulate the method of the
investigation in Ref.\ \cite{ONOB}.  We put $A_{tot}$ nucleons into
the potential wall with a large radius and give them the total energy
$E_{tot}$.  The confinement of nucleons inside the potential wall is
made by adding to the AMD Hamiltonian of Eq.\ (\ref{amdhamil}) the
following term
\begin{equation}
\frac k 2 \sum_{j=1}^{A_{tot}} f(|{\vec D}_j -{\vec D}_{CM}|)
\end{equation}
with
\begin{eqnarray}
&& f(x) = (x-a)^2 \theta (x-a), \nonumber \\
&& {\vec D}_j = \mathop{\rm Re} {\vec Z}_j/\sqrt \nu ,\quad
  {\vec D}_{CM} =
\frac 1 {A_{tot}} \sum_{j=1}^{A_{tot}} {\vec D}_j, \\
&& a = 12 {\rm fm},\quad   k = 5 {\rm MeV/fm^2}. \nonumber
\end{eqnarray}
Here $\theta(x)$ is the step function.  When we calculate the time
evolution of the system for a long time ($\sim 10^5$ fm/c), the system
is in a phase equilibrium of liquid nucleons and gas nucleons.  By
liquid nucleons we imply bound nucleons forming a nucleus.  The mass
number $A_{liq}$ of liquid nucleons changes with time and we select
the moments at which $A_{liq}$ takes a given value in order to get the
statistical properties of the nucleus with the given mass number.  We
calculate the temperature $T_g$ and the long-time average value of the
energy $E_{liq}$ of the liquid nucleons.  The temperature $T_g$, which
should be common to both phases, is calculated as the long-time
average value of $\tau$, where $(3/2)\tau$ is the kinetic energy (plus
the potential energy from the wall) per nucleon in the gas phase.
Since the gas phase is dilute, the effect of the Pauli principle is
neglected. When there exist some fragments in the gas phase, they are
excluded in the calculation of the temperature.  In the calculation of
the temperature $T_g$, i.e. the kinetic energy of gas nucleons, we
only use the central value of the wave packet $\mathop{\rm Im}{\vec
Z}_j$ in the momentum space.  It means that we treat gas nucleons as
point particles.  The treatment of gas nucleons as point particles is
consistent with the subtraction of the zero-point energies of nucleons
in the AMD Hamiltonian $\cal H$ of Eq.\ (\ref{amdhamil}).  It is to be
noted that for dilute gas in a large box there is no difference
between quantum and classical statistics.  The single-particle energy
spacing in the box with the size $2a$ is $(\hbar^2/2M)(\pi/2a)^2
\approx$ 0.35 MeV which is sufficiently small in our present problem.

In Ref.\ \cite{ONOB} we showed that the relation between $E_{liq}$ and
$T_g$ is very close to the linear curve $E_{liq}^*/A_{liq} = 3 T_g$
with $E_{liq}^* \equiv E_{liq} - E_{liq}({\rm ground\ state})$. It is
displayed again in Fig.\ 3. On the other hand, according to the proof
in Sec.\ III, the caloric curve between $E_{liq}$ and the temperature
$T_c$ of wave packet centroids of liquid nucleons should be very close
to the linear curve $E_{liq}^*/A_{liq} = 3 T_c$. Thus we obtain the
relation of $T_g = T_c$. This relation verifies the correctness of our
conjecture that in the ordinary AMD the emitted nucleons reflect not
the correct momentum distribution of nucleons but only the momentum
distribution of wave packet centroids inside the nucleus.

\begin{figure}
\ifx\epsfxsize\undefined\else
\begin{minipage}[b]{0.5\textwidth}
\epsfxsize\textwidth\epsfbox{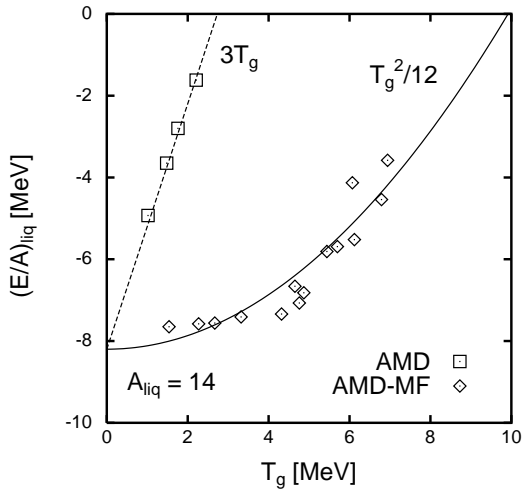}
\end{minipage}
\fi
\begin{minipage}[b]{0.48\textwidth}
\caption{
The statistical property of the excited nucleus calculated in the
phase equilibrium of liquid phase and gas phase. Results are shown for
the usual AMD (squares) and for AMD-MF (diamonds). Lines of
$(E^*/A)_{liq}=3T_g$ and $T_g^2/(12\,{\rm MeV})$ are drawn for the
comparison.}
\end{minipage}
\end{figure}

By using the calculated linear relation between $T_g$ and $E_{liq}$ we
can discuss conversely as follows.  We first regard the gas nucleons
as being a thermometer which measures the temperature $T_c$ of wave
packet centroids of liquid nucleons. Namely we regard that the
temperature $T_g$ is just the temperature $T_c$ of wave packet
centroids of liquid nucleons. It is justified because the interaction
between the thermometer and the liquid nucleons is described only by
the equation of motion of wave packet centroids.  Then, the calculated
linear relation between $T_g$ and $E_{liq}$ gives us a novel proof of
the classical statistics of wave packet centroids.  This proof is
independent of the proof given in Sec.\ III by the use of the
microcanonical ensemble technique.

Now we discuss the improvement of the description of the
nucleon-emission dynamics so that the high momentum component of the
wave packet can go out of the nucleus under the situation where the
wave packet centroid can not go out of the nucleus. For the
improvement, we have to treat the splitting of the wave packet into
high and low momentum parts, one going out of the nucleus and the
other remaining inside the nucleus.  It requires us an introduction of
a new stochastic process into AMD.  There can be various kinds of new
description of the nucleon emission.  From the arguments given above,
the essential point which any kind of new description should have is
that the nucleon emission through the splitting of a wave packet is
made according at least to the momentum distribution of the wave
packet. In Ref.\ \cite{ONOB} we presented a new description named
AMD-MF in which we apply the new description only when a nucleon is
near the nuclear surface and is going to be emitted from the nucleus.
For more detailed explanation of the AMD-MF the reader is referred to
Ref.\ \cite{ONOB}.

When we introduce a new description of the nucleon emission, we have
to check whether the new description ensures correctly the quantum
statistics or not for the nucleon-emission process. We discuss the
check of the statistics of the AMD-MF also by investigating the
equilibrated coexistent situation of gas and liquid nucleons.  This
investigation was made in Ref.\ \cite{ONOB} entirely in the same way
as discussed above for the check in the case of the usual AMD.  The
calculated relation between $T_g$ and $E_{liq}$ by the use of AMD-MF
was found to be approximated by the caloric curve $E_{liq}^*/A_{liq} =
T_g^2/(12 {\rm MeV})$ with $E_{liq}^* \equiv E_{liq} - E_{liq}({\rm
ground\ state})$ which is similar to the empirical Fermi gas caloric
curve.  It is displayed again in Fig.\ 3. This result strongly
supports that the temperature $T_g$ can be regarded as being equal to
the temperature $T$ of wave packets of liquid nucleons.  We can say
that the new description of the nucleon emission in AMD-MF ensures
proper quantum statistics for the nucleon-emission process.

Here also we can discuss conversely by using the calculated result of
$E_{liq}^*/A_{liq} \approx T_g^2/(12 {\rm MeV})$.  We first regard the
gas nucleons as being a thermometer which measures the temperature $T$
of wave packets of liquid nucleons. Namely we regard that the
temperature $T_g$ is just the temperature $T$ of wave packets of
liquid nucleons.  The justification of this identification is due to
the proper treatment of the nucleon emission in AMD-MF.  In AMD-MF the
nucleon emission is made properly according to the momentum
distribution of wave packet. Thus we have a good reason to believe
that the energy distribution of the emitted nucleons reflects
adequately the energy distribution of liquid nucleons which is
composed of the convolution of the momentum spread of wave packet and
the momentum distribution of wave packet centroids.  Then, the
calculated caloric curve $E_{liq}^*/A_{liq} \approx T_g^2/(12 {\rm
MeV})$ gives us a novel proof of the quantum statistics of wave
packets of liquid nucleons in AMD.  This proof is not only independent
of the proof given in Sec.\ IV but also more convincing than it,
because the system treated here is a realistic nuclear system while
the systems treated in Sec.\ IV are simple model systems under
one-dimensional harmonic oscillator mean field.  It is to be stressed
here that the present proof elucidates most straightforwardly the fact
that the wave packet spread causes the transition from the classical
statistics of wave packet centroids to the quantum statistics of wave
packets.

Before closing this section, we comment on the treatment of emitted
nucleons as point particles in the case of the usual AMD.  As
mentioned already, this treatment is due to our AMD Hamiltonian ${\cal
H}$ of Eq.\ (\ref{amdhamil}) where the zero-point energies of nucleons
are subtracted. The subtraction of the zero-point energy means that
the zero-point energy is converted into the translational kinetic
energy $K^2/2M$ ,i.e., that the wave packet in momentum space is
forced to shrink into a plane wave or a point particle. If we do not
subtract the zero-point energy $3\hbar^2 \nu/2M$, we can not have
low-energy nucleons whose kinetic energy are smaller than $3\hbar^2
\nu/2M$.  It means that the temperature $T_g$ of gas nucleons can not
be lower than $\hbar^2 \nu/M$.  Moreover, it means that a nucleon can
not be emitted from the nucleus if the available energy of a nucleon
is smaller than $3\hbar^2 \nu/2M$.  Even when a nucleon can avoid this
spurious hindrance of nucleon emission, the emission rate is forced to
be spuriously reduced.  Now let us consider the usual AMD with the
subtraction of zero-point energies of emitted nucleons. Although the
subtraction of the zero-point energy favors the nucleon emission
compared to the case without subtraction, it still underestimates the
nucleon emission rate.  The reason is given just by what we discussed
in this section. Namely, in the usual AMD the nucleon emission due to
the high momentum tail of the nucleon wave packet is neglected. Here,
in relation with these discussions, we consider the treatment of the
width parameter $\nu$ of the wave packet as dynamical variables as in
FMD.  In such treatment, the result will become somewhat between two
cases with and without the zero-point energy subtraction.  Therefore
the satisfactory treatment of the nucleon emission will be impossible
as long as we require the whole wave packet should go out in the
process.  Finally we comment on the ad hoc prescription to put the
wave packet spread to the emitted nucleons in the usual AMD with the
zero-point energy subtraction.  This prescription may be regarded as
simulating the effect of the emission due to the high momentum tail of
the wave packet.  It surely improves the fitting of the calculated
momentum distribution of emitted nucleons to data in the high momentum
region. However this prescription has nothing to do with the emission
dynamics and for example the slow emission rate remains untouched.  If
we calculate the gas temperature $T_g$ with this ad hoc prescription,
the change of $T_g$ is simply an increase by $\hbar^2 \nu/M$ and we
have no lower temperature than $\hbar^2 \nu/M$, showing inadequacy of
this temperature.

\section{SUMMARIZING DISCUSSIONS}

We discussed in detail the statistical properties of AMD in general
situations including both processes without and with nucleon
emissions. The most important point of our discussion was to clarify
whether the AMD statistics is quantum mechanical or classical. The
basic background for this point of discussion was as follows: AMD is a
kind of wave packet molecular dynamics which has been developed in
various forms especially in nuclear physics in order to cope with
quantum mechanical features of the nuclear system.  Since the point
particle molecular dynamics used widely in other fields like molecular
physics and solid state physics is of classical nature, one tends to
think that the wave packet molecular dynamics is also of classical
nature.  Our basic motivation was to show in a clear way that this
understanding is wrong.  We stressed that the strongest suggestion
that the wave packet molecular dynamics cannot be simply classical
comes from the following fact.  Namely, the exact solution of the
time-dependent Schr\"odinger equation of the many-body harmonic
oscillator system is given by the wave function of the Gaussian wave
packet molecular dynamics.  This fact is well known for the system of
distinguishable particles. This fact is, however, also true for the
system of fermions as can be easily proved.

In order to discuss the statistical properties of wave packet
molecular dynamics, we have to clarify the roles of two ingredients of
wave packet molecular dynamics, i.e. wave packet centroids and wave
packet spread.  Needless to say, one cannot at all make light of the
role of wave packet spread in the discussion of statistical properties
of wave packet molecular dynamics.  For example, the momentum
distribution of the system can not be described without wave packet
spread.  In general, in wave packet molecular dynamics physical
quantities are not described simply by using only wave packet
centroids but by using the wave function including the wave packet
spread.

One of the basic points of this paper was to elucidate the fact that
the presence of two ingredients gives rise to a very characteristic
feature of the AMD statistics.  It is the coexistence of mutually
opposite two statistics in the AMD framework, one being the classical
statistics which wave packet centroids obey in their motion described
by the AMD equation of motion and the other being the quantum
statistics which wave packets obey in their motion described by AMD
wave function.  The coexistence of the opposite two statistics is also
true for the wave packet molecular dynamics of distinguishable
particles and is a very important and novel character of the wave
packet molecular dynamics which constitutes a decisive difference from
the point particle molecular dynamics.

The fact that the statistics of wave packet centroids in AMD is
classical in spite of the antisymmetrization was proved in two
independent ways in this paper.  Both of the proofs were different
from the proof of Ref.\ \cite{OHNISHI} by Ohnishi and Randrup and were
made in the case of realistic many-nucleon systems interacting with
effective two-nucleon force.  One was a rather direct proof by the use
of the microcanonical ensemble technique.  The other was a proof by
the use of a thermometer made of gas nucleons which measures the
temperature of the motion of wave packet centroids inside the nucleus.
The reason why the thermometer measures the temperature of the wave
packet centroids is because the interaction between the thermometer
and the nucleons inside the nucleus is governed only by the motion of
wave packet centroids without any effects of wave packet spread.

The fact that the AMD statistics of wave packets including effects of
wave packet spread is quantum mechanical was also given two
independent proofs in this paper.  Both of them are different from the
proof of Ref.\ \cite{SCHNACK} by Schnack and Feldmeier.  One was a
proof in the case of one-dimensional harmonic oscillator systems of
fermions and also of distinguishable particles.  We calculated the
occupation probability of the harmonic oscillator single-particle
level on the basis of the fact that wave packet centroids obey
classical statistics. We then showed that the calculated values of the
occupation probability were almost the same as the values given by the
quantum canonical ensemble average.  What was decisively important in
this proof was the recognition of the fact that the temperature of the
classical statistics of wave packet centroids is different from the
temperature which characterizes the quantum statistics of wave
packets.  A merit of this proof is that it clarifies the relation
between the classical statistics of wave packet centroids and the
quantum statistics of wave packets, or in other words, it clarifies
how the quantum statistics of wave packets emerges from the classical
statistics of wave packet centroids. The other proof was by the use of
a new thermometer made of gas nucleons which measures the temperature
of the motion of wave packets inside the nucleus.  The new thermometer
interacts with the nucleus now through a new mechanism of the nucleon
emission from the nucleus.  The nucleon emission from the nucleus is
made not by the ordinary AMD but by the modified AMD named AMD-MF
which was proposed in Ref.\ \cite{ONOB}.  In the AMD-MF, the nucleon
emission is governed not by the motion of the wave packet centroids
but by the momentum distribution due to the wave packet spread.  A
merit of this proof is that it clarifies most straightforwardly the
role of the wave packet spread for the transition from the classical
statistics of wave packet centroids to the quantum statistics of wave
packets.

The temperature which is relevant to the calculated observables in AMD
is of course not the temperature $T_c$ of wave packet centroids but
the temperature $T$ of wave packets.  The serious mistake of the work
of Ref.\ \cite{OHNISHI} by Ohnishi and Randrup is that they regarded
the temperature $T_c$ of wave packet centroids as being the
temperature relevant to the calculated observables.  Based on this
mistake they insisted a wrong opinion that one should correct the AMD
framework even for the harmonic oscillator many-body system for which
the AMD wave function gives the exact solution of the time-dependent
Schr\"odinger equation.

The conclusion of this paper about the character of the AMD statistics
can be said as follows: As far as the effects of the wave packet
spread are duely taken into account, the AMD statistics is quantum
mechanical.  For processes without nucleon emission, namely for the
motion confined inside the nucleus, the ordinary AMD takes due account
of the wave packet spread and the AMD statistics is quantum mechanical
without any need of essential modification.  However, for the
nucleon-emission process, the ordinary AMD does not duely take account
of the wave packet spread.  Namely in the ordinary AMD, the emitted
nucleons reflect not the momentum distribution of nucleons inside the
nucleus but that of wave packet centroids inside the nucleus.  Hence
the statistics of the ordinary AMD for the nucleon-emission process is
classical.  In order to recover the quantum statistics also for the
nucleon-emission process, the description of the nucleon-emission
process of the ordinary AMD should be modified so that the nucleon
emission is made according to the momentum distribution of nucleons
inside the nucleus which is the convolution of the momentum spread of
the wave packet and the momentum distribution of wave packet
centroids.  The AMD-MF method proposed in Ref.\ \cite{ONOB} is one of
such modified versions of AMD and it was verified that that the
statistics of the AMD-MF for the nucleon-emission process is actually
quantum mechanical.

What is to be noticed in the above discussion is that the reason of
the classical statistics of the usual AMD for the nucleon-emission
process is originating not from any complex mechanism but from a
simple mechanism of the single-particle motion of emission.  Thus we
have been able to recover the proper quantum statistics for the
nucleon-emission process by a modification of the single-particle
dynamics.  In the modification, a proper incorporation of the wave
packet spread into the dynamical process should be made. Here,
however, this modification should not be confused with the treatment
of the width parameters of wave packets as time-dependent dynamical
variables.

The modification of the AMD by the AMD-MF can be generalized so that
we can improve the description of the cluster emission process and
furthermore that of the nucleon-transfer process between the
projectile and the target in addition to that of the nucleon-emission
process.  We have already formulated such extension and have found
that fragmentation processes are largely influenced by the extended
modification.  We will discuss these in other papers.

\end{document}